\documentclass[12pt,cls,onecolumn]{IEEEtran}
\usepackage{subfigure,cite,graphicx,psfig,amsmath,amssymb,mathrsfs,epsf}

\begin{document}

\title{On Capacity Scaling of Underwater Networks: An Information-Theoretic Perspective}
\author{\large Won-Yong Shin, \emph{Member}, \emph{IEEE}, Daniel E. Lucani, \emph{Student Member}, \emph{IEEE}, \\Muriel M{\'e}dard,
\emph{Fellow}, \emph{IEEE}, Milica Stojanovic,
\emph{Fellow}, \emph{IEEE},\\and Vahid Tarokh, \emph{Fellow}, \emph{IEEE} \\
\thanks{This work was supported in part by the National Science Foundation under grants No. 0520075, 0831728 and
CNS-0627021, and by the ONR MURI grant No. N00014-07-1-0738,
subcontract \# 060786 issued by BAE Systems National Security
Solutions, Inc., and supported by the Defense Advanced Research
Projects Agency (DARPA) and the Space and Naval Warfare System
Center (SPAWARSYSCEN), San Diego under Contract No. N66001-06-C-2020
(CBMANET), by the National Science Foundation under grant No.
0831728, and by the ONR grant No. N00014-09-1-0700. The material in
this paper is to be presented in part at the IEEE International
Symposium on Information Theory, Austin, TX, June 2010.}
\thanks{W.-Y. Shin and V. Tarokh are with the School of Engineering and Applied Sciences, Harvard
University, Cambridge, MA 02138 USA (E-mail:\{wyshin,
vahid\}@seas.harvard.edu).}
\thanks{D. E. Lucani and M. M{\'e}dard are with the Research Laboratory of Electronics, Massachusetts Institute of Technology,
Cambridge, MA 02139 USA (E-mail: \{dlucani, medard\}@mit.edu).}
\thanks{M. Stojanovic is with the ECE Department, Northeastern University, Boston, MA 02115 USA (E-mail: millitsa@mit.edu).}
        } \maketitle


\markboth{Submitted to IEEE Transactions on Information Theory}
{Shin {\em et al.}: On Capacity Scaling of Underwater Networks: An
Information-Theoretic Perspective}


\newtheorem{definition}{Definition}
\newtheorem{theorem}{Theorem}
\newtheorem{lemma}{Lemma}
\newtheorem{example}{Example}
\newtheorem{corollary}{Corollary}
\newtheorem{proposition}{Proposition}
\newtheorem{conjecture}{Conjecture}
\newtheorem{remark}{Remark}

\def \diag{\operatornamewithlimits{diag}}
\def \min{\operatornamewithlimits{min}}
\def \max{\operatornamewithlimits{max}}
\def \log{\operatorname{log}}
\def \max{\operatorname{max}}
\def \rank{\operatorname{rank}}
\def \out{\operatorname{out}}
\def \exp{\operatorname{exp}}
\def \arg{\operatorname{arg}}
\def \E{\operatorname{E}}
\def \tr{\operatorname{tr}}
\def \SNR{\operatorname{SNR}}
\def \dB{\operatorname{dB}}
\def \ln{\operatorname{ln}}

\def \be {\begin{eqnarray}}
\def \ee {\end{eqnarray}}
\def \ben {\begin{eqnarray*}}
\def \een {\end{eqnarray*}}

\begin{abstract}
Capacity scaling laws are analyzed in an underwater acoustic network
with $n$ regularly located nodes on a square. A narrow-band model is
assumed where the carrier frequency is allowed to scale as a
function of $n$. In the network, we characterize an attenuation
parameter that depends on the frequency scaling as well as the
transmission distance. A cut-set upper bound on the throughput
scaling is then derived in extended networks. Our result indicates
that the upper bound is inversely proportional to the attenuation
parameter, thus resulting in a highly power-limited network.
Interestingly, it is seen that unlike the case of wireless radio
networks, our upper bound is intrinsically related to the
attenuation parameter but not the spreading factor. Furthermore, we
describe an achievable scheme based on the simple nearest-neighbor
multi-hop (MH) transmission. It is shown under extended networks
that the MH scheme is order-optimal as the attenuation parameter
scales exponentially with $\sqrt{n}$ (or faster). Finally, these
scaling results are extended to a random network realization.
\end{abstract}


\newpage


\section{Introduction}

A pioneering work of~\cite{GuptaKumar:00}, introduced by Gupta and
Kumar, characterized the sum throughput scaling in a large wireless
radio network. They showed that the total throughput scales as
$\Theta(\sqrt{n/\log n})$ when a multi-hop (MH) routing strategy is
used for $n$ source--destination (S--D) pairs randomly distributed
in a unit area.\footnote{We use the following notations: i)
$f(x)=O(g(x))$ means that there exist constants $C$ and $c$ such
that $f(x)\le Cg(x)$ for all $x>c$. ii) $f(x)=o(g(x))$ means that
$\underset{x\rightarrow\infty}\lim\frac{f(x)}{g(x)}=0$. iii)
$f(x)=\Omega(g(x))$ if $g(x)=O(f(x))$. iv) $f(x)=\omega(g(x))$ if
$g(x)=o(f(x))$. v) $f(x)=\Theta(g(x))$ if $f(x)=O(g(x))$ and
$g(x)=O(f(x))$~\cite{Knuth:76}.} MH schemes are then further
developed and analyzed
in~\cite{GuptaKumar:03,FranceschettiDouseTseThiran:07,XueXieKumar:05,ElGamalMammenPrabhakarShah:06,ElGamalMammen:06,NebatCruzBhardwaj:09,
ShinChungLee:09}, while their throughput per S--D pair scales far
slower than $\Theta(1)$. Recent
results~\cite{OzgurLevequeTse:07,NiesenGuptaShah:09} have shown that
an almost linear throughput in the radio network, i.e.
$\Theta(n^{1-\epsilon})$ for an arbitrarily small $\epsilon>0$,
which is the best we can hope for, is achievable by using a
hierarchical cooperation strategy. Besides the schemes
in~\cite{OzgurLevequeTse:07,NiesenGuptaShah:09}, there have been
other studies to improve the throughput of wireless radio networks
up to a linear scaling in a variety of network scenarios by using
novel techniques such as networks with node
mobility~\cite{GrossglauserTse:02}, interference
alignment~\cite{CadambeJafar:07}, and infrastructure
support~\cite{ZemlianovVeciana:05}.

Along with the studies in terrestrial radio networks, the interest
in study of underwater networks has been growing with recent
advances in acoustic communication
technology~\cite{PartanKuroseLevine:06,Stojanovic:07,LucaniMedardStojanovicJSAC:08,LucaniStojanovicMedardOcean:08}.
In underwater acoustic communication systems, both bandwidth and
power are severely limited due to the exponential (rather than
polynomial) path-loss attenuation with propagation distance and even
frequency-dependent attenuation. This is a main feature that
distinguishes underwater systems from wireless radio links. Hence,
the system throughput is affected by not only the transmission
distance but also the useful bandwidth. Based on these
characteristics, network coding
schemes~\cite{GuoXieCuiWang:06,LucaniMedardStojanovic:07,LucaniMedardStojanovicJSAC:08}
have been presented for underwater acoustic channels, while network
coding showed better performance than MH routing in terms of
reducing transmission power. MH networking has further been
investigated in other simple but realistic network conditions that
take into account the practical issues of coding and
delay~\cite{CarbonelliMitra:06,ZhangStojanovicMitra:08}.

One natural question is what are the fundamental capabilities of
underwater networks in supporting a multiplicity of nodes that wish
to communicate concurrently with each other, i.e., multiple S--D
pairs, over an acoustic channel. To answer this question, the
throughput scaling for underwater networks was first
studied~\cite{LucaniMedardStojanovic:08}, where $n$ nodes were
arbitrarily located in a planar disk of unit
area~\cite{GuptaKumar:00} and the carrier frequency was set to a
constant independent of $n$. That work showed an upper bound on the
throughput of each node based on the physical model assumption
in~\cite{GuptaKumar:00}. This upper bound scales as
$n^{-1/\alpha}e^{-W_0(\Theta(n^{-1/\alpha}))}$, where $\alpha$
corresponds to the spreading factor of the underwater channel, and
$W_0$ represents the branch zero of the Lambert
function~\cite{Chapeau-BlondeauMonir:02}.\footnote{The Lambert W
function is defined to be the inverse of the function
$z=W(z)e^{W(z)}$ and the branch satisfying $W(z)\ge-1$ is denoted by
$W_0(z)$.} Since the spreading factor typically has values in the
range $1 \leq \alpha \leq 2$~\cite{LucaniMedardStojanovic:08}, the
throughput per node decreases almost as $O(n^{-1/\alpha})$ for large
enough $n$, which is considerably faster than the $\Theta(\sqrt{n})$
scaling characterized for wireless radio
settings~\cite{GuptaKumar:00}.

In this paper, a capacity scaling law for underwater networks is
analyzed in extended
networks~\cite{XieKumar:04,JovicicViswanathKulkarni:04,XueXieKumar:05,FranceschettiDouseTseThiran:07,OzgurLevequeTse:07}
of unit node density. Especially, we are interested in the case
where the carrier frequency scales as a certain function of $n$ in a
narrow-band model. Such an assumption changes the scaling behavior
significantly due to the attenuation characteristics. We aim to
study both an information-theoretic upper bound and achievable rate
scaling while allowing the frequency scaling with $n$.

We explicitly characterize an {\em attenuation parameter} that
depends on the transmission distance and also on the carrier
frequency, and then identify fundamental operating regimes depending
on the parameter. For networks with $n$ regularly distributed nodes,
we derive an upper bound on the total throughput scaling using the
cut-set bound. Our upper bound is based on the characteristics of
power-limited regimes shown
in~\cite{OzgurLevequeTse:07,OzgurJohariTseLeveque:10}. In extended
networks, it is shown that the upper bound is inversely proportional
to the attenuation parameter. This leads to a highly power-limited
network for all the operating regimes, where power consumption is
important in determining performance. Interestingly, it is seen that
unlike the case of wireless radio networks, our upper bound heavily
depends on the attenuation parameter but not on the spreading factor
(corresponding to the path-loss exponent in wireless networks). In
addition, to constructively show our achievability result for
extended regular networks, we describe the conventional
nearest-neighbor MH transmission~\cite{GuptaKumar:00} with a slight
modification, and analyze its achievable throughput. It is shown
under extended networks that the achievable rate based on the MH
routing scheme matches the upper bound within a factor of $n$ with
arbitrarily small exponent as long as the attenuation parameter
increases exponentially with respect to $\sqrt{n}$ (or faster).
Furthermore, a random network scenario is also discussed in this
work. It is shown under extended random networks that the
conventional MH-based achievable scheme is not order-optimal for any
operating regimes.

The rest of this paper is organized as follows.
Section~\ref{SEC:System} describes our system and channel models. In
Section~\ref{SEC:Upper}, the cut-set upper bound on the throughput
is derived. In Section~\ref{SEC:Achievability}, achievable
throughput scaling is analyzed. These results are extended to the
random network case in Section~\ref{SEC:Random}. Finally, Section
\ref{SEC:Conc} summarizes the paper with some concluding remarks.

Throughout this paper the superscript $H$, $[\cdot]_{ki}$, and
$\|\cdot\|_2$ denote the conjugate transpose, the $(k, i)$-th
element, and the largest singular value, respectively, of a matrix.
${\bf I}_n$ is the identity matrix of size $n\times n$, $\tr(\cdot)$
is the trace, and $\det(\cdot)$ is the determinant. $\mathbb{C}$ is
the field of complex numbers and $E[\cdot]$ is the expectation.
Unless otherwise stated, all logarithms are assumed to be to the
base 2.


\section{System and Channel Models} \label{SEC:System}

We consider a two-dimensional underwater network that consists of
$n$ nodes on a square with unit node density such that two
neighboring nodes are 1 unit of distance apart from each other in an
extended network\footnote{A dense
network~\cite{GuptaKumar:00,ElGamalMammenPrabhakarShah:06,OzgurLevequeTse:07}
of unit area can also be considered as another fundamental network
model, which will not be shown in this work. We remark that there
exists either a bandwidth or power limitation (or both) according to
the path-loss attenuation regimes in dense networks.}, i.e., a
regular network~\cite{XieKumar:04,JovicicViswanathKulkarni:04}. We
randomly pick a matching of S--D pairs, so that each node is the
destination of exactly one source. We assume frequency-flat channel
of bandwidth $W$ Hz around carrier frequency $f$, which satisfies
$f\gg W$, i.e., narrow-band model. This is a highly simplified
model, but nonetheless one that suffices to demonstrate the
fundamental mechanisms that govern capacity scaling. Assuming that
all the nodes have perfectly directional transmissions, we also
disregard multipath propagation, and simply focus on a line-of-sight
channel between each pair of nodes used
in~\cite{OzgurLevequeTse:07,NiesenGuptaShah:09,OzgurJohariTseLeveque:10}.
Each node has an average transmit power constraint $P$ (constant),
and we assume that the channel state information (CSI) is available
at all receivers, but not at the transmitters. It is assumed that
each node transmits at a rate $T(n)/n$, where $T(n)$ denotes the
total throughput of the network.

Now let us turn to channel modeling. An underwater acoustic channel
is characterized by an attenuation that depends on both the distance
$r_{ki}$ between nodes $i$ and $k$ ($i, k\in\{1,\cdots,n\}$) and the
signal frequency $f$, and is given by
\begin{equation}   \label{EQ:af}
A(r_{ki},f)=c_0 r_{ki}^{\alpha}a(f)^{r_{ki}}
\end{equation}
for some constant $c_0>0$ independent of $n$, where $\alpha$ is the
spreading factor and $a(f)>1$ is the absorption
coefficient~\cite{Stojanovic:07}. For analytical convenience, we
assume that the spreading factor $\alpha$ does not change throughout
the network, i.e., that it is the same from short to long range
transmissions, as in wireless radio
networks~\cite{GuptaKumar:00,FranceschettiDouseTseThiran:07,OzgurLevequeTse:07}.
The spreading factor describes the geometry of propagation and is
typically $1\leq \alpha \leq 2$---its commonly used values are
$\alpha=1$ for cylindrical spreading, $\alpha=2$ for spherical
spreading, and $\alpha=1.5$ for the so-called practical spreading.
Note that existing models of wireless networks typically correspond
to the case for which $a(f)=1$ (or a positive constant independent
of $n$) and $\alpha>2$.\footnote{The counterpart of $\alpha$ in
wireless radio channels is the path-loss exponent.}

A common empirical model gives $a(f)$ in dB/km for $f$ in kHz
as~\cite{BerkhovskikhLysanov:82,Stojanovic:07}:
\begin{equation} \label{EQ:afeq}
10\log a(f)=a_0+a_1
f^2+a_2\frac{f^2}{b_1+f^2}+a_3\frac{f^2}{b_2+f^2},
\end{equation}
where $\{a_0,\cdots,a_3,b_1,b_2\}$ are some positive constants
independent of $n$. As mentioned earlier, we will allow the carrier
frequency $f$ to scale with $n$. Especially, we consider the case
where the frequency scales at arbitrarily increasing rates relative
to $n$, which enables us to really capture the dependence on the
frequency in performance.\footnote{Otherwise, the attenuation
parameter $a(f)$ scales as $\Theta(1)$ from (\ref{EQ:afeq}), which
is not a matter of interest in this work.} The absorption $a(f)$ is
then an increasing function of $f$ such that
\begin{equation} \label{EQ:afTheta}
a(f)=\Theta\left(e^{c_1 f^2}\right)
\end{equation}
with respect to $f$ for some constant $c_1>0$ independent of $n$.

The noise $n_i$ at node $i\in\{1,\cdots,n\}$ in an acoustic channel
can be modeled through four basic sources: turbulence, shipping,
waves, and thermal noise~\cite{Stojanovic:07}. We assume that $n_i$
is the circularly symmetric complex additive colored Gaussian noise
with zero mean and power spectral density (psd) $N(f)$, and thus the
noise is frequency-dependent. The overall psd of four sources decays
linearly on the logarithmic scale in the frequency region 100 Hz --
100 kHz, which is the operating regime used by the majority of
acoustic systems, and thus is approximately given
by~\cite{Coates:89,Stojanovic:07}
\begin{equation} \label{EQ:Nfeq}
\log N(f)= a_4-a_5\log f
\end{equation}
for some positive constants $a_4$ and $a_5$ independent of
$n$.\footnote{Note that in our operating frequencies, $a_5=1.8$ is
commonly used for the above approximation~\cite{Stojanovic:07}.}
This means that $N(f)=O(1)$ since
\begin{equation} \label{EQ:NfTheta}
N(f)=\Theta\left(\frac{1}{f^{a_5}}\right)
\end{equation}
in terms of $f$ increasing with $n$. From (\ref{EQ:afTheta}) and
(\ref{EQ:NfTheta}), we may then have the following relationship
between the absorption $a(f)$ and the noise psd $N(f)$:
\begin{equation} \label{EQ:Nfaf}
N(f)=\Theta\left(\frac{1}{(\log a(f))^{a_5/2}}\right).
\end{equation}

From the narrow-band assumption, the received signal $y_k$ at node
$k \in \{1,\cdots,n\}$ at a given time instance is given by
\begin{equation}
y_k=\sum_{i\in I}h_{ki}x_i+n_k, \label{EQ:signal}
\end{equation}
where
\begin{equation}
h_{ki}=\frac{e^{j\theta_{ki}}}{\sqrt{A(r_{ki},f)}} \label{EQ:hki}
\end{equation}
represents the complex channel between nodes $i$ and $k$,
$x_i\in\mathbb{C}$ is the signal transmitted by node $i$, and $I
\subset \{1,\cdots,n\}$ is the set of simultaneously transmitting
nodes. The random phases $e^{j\theta_{ki}}$ are uniformly
distributed over $[0, 2\pi)$ and independent for different $i$, $k$,
and time. We thus assume a narrow-band time-varying channel, whose
gain changes to a new independent value for every symbol. Note that
this random phase model is based on a far-field
assumption~\cite{OzgurLevequeTse:07,NiesenGuptaShah:09,OzgurJohariTseLeveque:10},\footnote{In~\cite{FranceschettiMiglioreMinero:09},
instead of simply taking the far-field assumption, the physical
limit of wireless radio networks has been studied under certain
conditions on scattering elements. Further investigation is also
required to see whether this assumption is valid for underwater
networks of unit node density in the limit of large number $n$ of
nodes.} which is valid if the wavelength is sufficiently smaller
than the minimum node separation.

Based on the above channel characteristics, operating regimes of the
network are identified according to the following physical
parameters: the absorption $a(f)$ and the noise psd $N(f)$ which are
exploited here by choosing the frequency $f$ based on the number $n$
of nodes. In other words, if the relationship between $f$ and $n$ is
specified, then $a(f)$ and $N(f)$ can be given by a certain scaling
function of $n$ from (\ref{EQ:afTheta}) and (\ref{EQ:NfTheta}),
respectively.


\section{Cut-set Upper Bound} \label{SEC:Upper}

To access the fundamental limits of an underwater network, a cut-set
upper bound on the total throughput scaling is analyzed from an
information-theoretic perspective~\cite{CoverThomas:91}.
Specifically, an upper bound based on the power transfer
argument~\cite{OzgurLevequeTse:07,OzgurJohariTseLeveque:10} is
established for extended networks. Note, however, that the present
problem is not equivalent to the conventional extended network
framework~\cite{OzgurLevequeTse:07} due to different channel
characteristics. Our interest is particularly in the operating
regimes for which the upper bound is tight.

\begin{figure}[t!]
  \begin{center}
  \leavevmode \epsfxsize=0.40\textwidth   
  \leavevmode 
  \epsffile{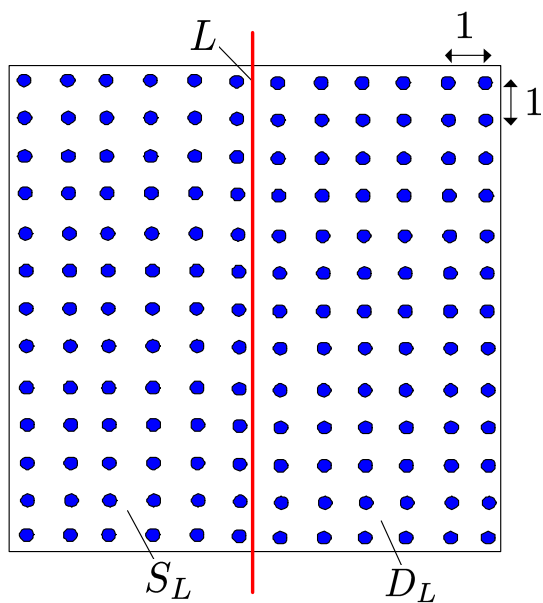}
  \caption{The cut $L$ in a two-dimensional extended regular network. $S_L$ and $D_L$ represent the sets of source and destination nodes, respectively.}
  \label{FIG:cut1}
  \end{center}
\end{figure}

Consider a given cut $L$ dividing the network area into two equal
halves as in~\cite{OzgurLevequeTse:07,OzgurJohariTseLeveque:10} (see
Fig.~\ref{FIG:cut1}). Let $S_L$ and $D_L$ denote the sets of sources
and destinations, respectively, for the cut $L$ in the network. More
precisely, under $L$, source nodes $S_L$ are on the left, while all
nodes on the right are destinations $D_L$. In this case, we have an
$\Theta(n)\times \Theta(n)$ multiple-input multiple-output (MIMO)
channel between the two sets of nodes separated by the cut.

In an extended network, we take into account an approach based on
the amount of power transferred across the network according to
different operating regimes, i.e., path-loss attenuation regimes. As
pointed out in~\cite{OzgurLevequeTse:07,OzgurJohariTseLeveque:10},
the information transfer from $S_L$ to $D_L$ is highly power-limited
since all the nodes in the set $D_L$ are ill-connected to the
left-half network in terms of power. This implies that the
information transfer is bounded by the total received power
transfer, rather than the cardinality of the set $D_L$. For the cut
$L$, the total throughput $T(n)$ for sources on the left is bounded
by the (ergodic) capacity of the MIMO channel between $S_L$ and
$D_L$ under time-varying channel assumption, and thus is given by
\begin{eqnarray} \label{EQ:Tn12}
T(n)\le \underset{{\bf Q}_L\ge0}\max E\left[\log\det\left({\bf
I}_{\Theta(n)}+\frac{1}{N(f)}{\bf H}_L{\bf Q}_L{\bf
H}_L^{H}\right)\right],
\end{eqnarray}
where ${\bf H}_L$ is the matrix with entries $[{\bf H}_L]_{ki}$ for
$i\in S_L, k\in D_L$, and ${\bf Q}_L\in
\mathbb{C}^{\Theta(n)\times\Theta(n)}$ is the positive semi-definite
input signal covariance matrix whose $k$-th diagonal element
satisfies $[{\bf Q}_L]_{kk}\le P$ for $k\in S_L$.

The relationship (\ref{EQ:Tn12}) will be further specified in
Theorem~\ref{TH:Cutset}. Before that, we first apply the techniques
of~\cite{OzgurLevequeTse:07,ShinJeonDevroyeVuChungLeeTarokh:08} to
obtain the total power transfer of the set $D_{L}$. These techniques
involve the relaxation of the individual power constraints to a
total weighted power constraint, where the weight assigned to each
source corresponds to the total received power on the other side of
the cut. To be more specific, each column $i$ of the matrix ${\bf
H}_{L}$ is normalized by the square root of the total received power
on the other side of the cut from source $i\in S_L$. From
(\ref{EQ:af}) and (\ref{EQ:hki}), the total power $P_{L}^{(i)}$
received from the signal sent by the source $i$ is given by
\begin{equation} \label{EQ:PL}
P_{L}^{(i)}=Pd_{L}^{(i)},
\end{equation}
where
\begin{eqnarray} \label{EQ:dL}
d_{L}^{(i)}=\frac{1}{c_0}\underset{k\in D_{L}}\sum
r_{ki}^{-\alpha}a(f)^{-r_{ki}}
\end{eqnarray}
for some constant $c_0>0$ independent of $n$. For convenience, we
now index the node positions such that the source and destination
nodes under the cut $L$ are located at positions $(-i_x+1,i_y)$ and
$(k_x,k_y)$, respectively, for $i_x, k_x=1,\cdots,\sqrt{n}/2$ and
$i_y, k_y=1,\cdots,\sqrt{n}$. The scaling result of $d_{L}^{(i)}$
defined in (\ref{EQ:dL}) can then be derived as follows.

\begin{lemma} \label{LEM:dL}
In an extended network, the term $d_{L}^{(i)}$ in (\ref{EQ:dL}) is
\begin{eqnarray} \label{EQ:dLupper}
d_{L}^{(i)}=\Theta\left(i_x^{1-\alpha}a(f)^{-i_x}\right),
\end{eqnarray}
where $-i_x+1$ represents the horizontal coordinate of node $i\in
S_L$ for $i_x=1,\cdots,\sqrt{n}/2$.
\end{lemma}

The proof of this lemma is presented in Appendix~\ref{PF:dL}. The
expression (\ref{EQ:Tn12}) is then rewritten as
\begin{equation} \label{EQ:logdetL2}
\underset{\tilde{\bf Q}_L\ge0}\max E\left[\log\det\left({\bf
I}_{\Theta(n)}+\frac{1}{N(f)}{\bf F}_{L}\tilde{\bf Q}_L{\bf
F}_{L}^{H}\right)\right],
\end{equation}
where ${\bf F}_{L}$ is the matrix with entries $[{\bf
F}_{L}]_{ki}=\frac{1}{\sqrt{d_{L}^{(i)}}}[{\bf H}_{L}]_{ki}$, which
are obtained from (\ref{EQ:dL}), for $i\in S_L, k\in D_{L}$. Here,
$\tilde{\bf Q}_L$ is the matrix satisfying
\begin{equation} \label{EQ:tildeQ}
\left[\tilde{\bf
Q}_L\right]_{ki}=\sqrt{d_{L}^{(k)}d_{L}^{(i)}}\left[{\bf
Q}_L\right]_{ki}, \nonumber
\end{equation}
which means that $\tr(\tilde{\bf Q}_L)\le\sum_{i\in S_L}P_{L}^{(i)}$
(equal to the sum of the total power received from each source).

We next examine the behavior of the largest singular value for the
normalized channel matrix ${\bf F}_{L}$, and then show how much it
affects an upper bound on (\ref{EQ:logdetL2}). We first address the
case where ${\bf F}_{L}$ is well-conditioned according to the
attenuation parameter $a(f)$.

\begin{lemma} \label{LEM:singular}
Let ${\bf F}_{L}$ denote the normalized channel matrix defined by
the expression (\ref{EQ:logdetL2}). Under the attenuation regimes
$a(f)=\Omega\left((1+\epsilon_0)^{\sqrt{n}}\right)$ for an
arbitrarily small $\epsilon_0>0$, we have that
\begin{equation} \label{EQ:EFL2}
E\left[\left\|{\bf F}_{L}\right\|_2^2\right]\le c_2\log n
\end{equation}
for some constant $c_2>0$ independent of $n$.
\end{lemma}

The proof of this lemma is presented in Appendix~\ref{PF:singular}.
Note that the matrix ${\bf F}_{L}$ is well-conditioned as $a(f)$
scales exponentially with respect to $\sqrt{n}$ (or faster).
Otherwise, i.e., if $a(f)=o\left((1+\epsilon_0)^{\sqrt{n}}\right)$,
the largest singular value of ${\bf F}_{L}$ scales as a polynomial
factor of $n$, thus resulting in a loose upper bound on the total
throughput. Using Lemma~\ref{LEM:singular}, we obtain the following
result.

\begin{lemma} \label{LEM:powerterm}
Under $a(f)=\Omega\left((1+\epsilon_0)^{\sqrt{n}}\right)$, the term
(\ref{EQ:logdetL2}) is upper-bounded by
\begin{equation} \label{EQ:PL2i}
\frac{n^{\epsilon}}{N(f)}\sum_{i\in S_L}P_{L}^{(i)}
\end{equation}
for arbitrarily small positive constants $\epsilon_0$ and
$\epsilon$, where $P_{L}^{(i)}$ is given by (\ref{EQ:PL}).
\end{lemma}

\begin{proof}
Equation (\ref{EQ:logdetL2}) is bounded by
\begin{eqnarray} \label{EQ:eventFL2}
&&\underset{\tilde{\bf Q}_L\ge0}\max E\left[\log\det\left({\bf
I}_{\Theta(n)}+\frac{1}{N(f)}{\bf F}_{L}\tilde{\bf Q}_L{\bf
F}_{L}^{H}\right)1_{\mathcal{E}_{{\bf F}_{L}}}\right] \nonumber\\
+\!\!\!\!\!\!\!&&\underset{\tilde{\bf Q}_L\ge0}\max
E\left[\log\det\left({\bf I}_{\Theta(n)}+\frac{1}{N(f)}{\bf
F}_{L}\tilde{\bf Q}_L{\bf F}_{L}^{H}\right)1_{\mathcal{E}_{{\bf
F}_{L}}^{c}}\right].
\end{eqnarray}
Here, the event $\mathcal{E}_{{\bf F}_{L}}$ refers to the case where
the channel matrix ${\bf F}_{L}$ is accidently ill-conditioned and
is given by
\begin{equation}
\mathcal{E}_{{\bf F}_{L}}=\left\{\left\|{{\bf
F}_{L}}\right\|_2^2>n^{\epsilon}\right\} \nonumber
\end{equation}
for an arbitrarily small constant $\epsilon>0$. Suppose that
$a(f)=\Omega\left((1+\epsilon_0)^{\sqrt{n}}\right)$. Then, by using
the result of Lemma~\ref{LEM:singular} and applying the proof
technique similar to that in Section V of~\cite{OzgurLevequeTse:07},
it is possible to prove that the first term in (\ref{EQ:eventFL2})
decays polynomially to zero with arbitrary exponent as $n$ tends to
infinity, and for the second term in (\ref{EQ:eventFL2}), it follows
that
\begin{eqnarray}
\underset{\tilde{\bf Q}_L\ge0}\max
\!\!\!\!\!\!\!\!\!&&E\left[\log\det\left({\bf
I}_{\Theta(n)}+\frac{1}{N(f)}{\bf F}_{L}\tilde{\bf Q}_L{\bf
F}_{L}^{H}\right)1_{\mathcal{E}_{{\bf F}_{L}}^{c}}\right] \nonumber\\
&& \le\underset{\tilde{\bf Q}_L\ge0}\max
E\left[\frac{1}{N(f)}\left\|{\bf
F}_{L}\right\|_2^2\tr\left(\tilde{\bf
Q}_L\right)1_{\mathcal{E}_{{\bf F}_{L}}^{c}}\right] \nonumber\\ &&
\le \frac{c_2\log n}{N(f)} \underset{\tilde{\bf Q}_L\ge0}\max
\tr\left(\tilde{\bf Q}_L\right) \nonumber\\ && \le
\frac{n^{\epsilon}}{N(f)}\sum_{i\in S_L}P_{L}^{(i)} \nonumber
\end{eqnarray}
for some constant $c_2>0$ independent of $n$, where the second
inequality holds by Lemma~\ref{LEM:singular}. This completes the
proof of this lemma.
\end{proof}

Note that (\ref{EQ:PL2i}) represents the total amount of received
signal-to-noise ratio from the set $S_L$ of sources to the set $D_L$
of destinations for a given cut $L$. We are now ready to show the
cut-set upper bound in extended networks.

\begin{theorem} \label{TH:Cutset}
For an underwater regular network of unit node density, where the
absorption coefficient $a(f)$ scales as
$\Omega\left((1+\epsilon_0)^{\sqrt{n}}\right)$ for an arbitrarily
small $\epsilon_0>0$, the total throughput $T(n)$ is upper-bounded
by
\begin{eqnarray} \label{EQ:Tnupper}
T(n)\le \frac{c_3n^{1/2+\epsilon}}{a(f)N(f)},
\end{eqnarray}
where $c_3>0$ is some constant independent of $n$ and $\epsilon>0$
is an arbitrarily small constant.
\end{theorem}

\begin{proof}
Suppose that $a(f)=\Omega\left((1+\epsilon_0)^{\sqrt{n}}\right)$.
Then from Lemmas~\ref{LEM:dL} and~\ref{LEM:powerterm}, we obtain the
following upper bound on the total throughput $T(n)$:
\begin{eqnarray}
T(n)\!\!\!\!\!\!\!&&\le \frac{n^{\epsilon}}{N(f)}\underset{i\in
S_L}\sum Pd_{L}^{(i)} \nonumber\\ &&\le
\frac{Pn^{\epsilon}}{N(f)}\sum_{i_x=1}^{\sqrt{n}/2}\sum_{i_y=1}^{\sqrt{n}}
d_{L}^{(i)} \nonumber\\ &&\le
\frac{c_4Pn^{1/2+\epsilon}}{N(f)}\sum_{i_x=1}^{\sqrt{n}/2}\frac{1}{i_x^{\alpha-1}a(f)^{i_x}}
\nonumber\\ &&\le
\frac{c_4Pn^{1/2+\epsilon}}{N(f)}\sum_{i_x=1}^{\sqrt{n}/2}\frac{1}{a(f)^{i_x}}
\nonumber\\ &&\le \frac{c_5Pn^{1/2+\epsilon}}{a(f)N(f)}, \nonumber
\end{eqnarray}
where $c_4$ and $c_5$ are some positive constants independent of
$n$, which is equal to (\ref{EQ:Tnupper}). This completes the proof
of the theorem.
\end{proof}

Note that this upper bound is expressed as a function of the
absorption $a(f)$ and the noise psd $N(f)$ while an upper bound for
wireless radio networks depends only on the constant value
$\alpha$~\cite{OzgurLevequeTse:07}. We remark that when
$a(f)=o\left((1+\epsilon_0)^{\sqrt{n}}\right)$, the upper bound
becomes boosted by a certain polynomial factor of $n$ (up to
$O(\sqrt{n})$) compared to the case shown in
(\ref{EQ:Tnupper}).\footnote{This statement could be rigorously
proved by following the steps similar to those shown in
Lemmas~\ref{LEM:singular} and~\ref{LEM:powerterm}, even if the
details are not shown in this paper.} In addition, using
(\ref{EQ:Nfaf}) in (\ref{EQ:Tnupper}) results in
\begin{equation}
T(n)=O\left(\frac{\left(\log
a(f)\right)^{a_5/2}n^{1/2+\epsilon}}{a(f)}\right) \nonumber
\end{equation}
for some constant $a_5>0$ shown in (\ref{EQ:Nfeq}). Finally, another
expression for the condition in which the upper bound in
(\ref{EQ:Tnupper}) holds is shown as follows.

\begin{remark}
We examine the relationship between the carrier frequency $f$ and
the number $n$ of nodes such that our upper bound holds. By using
(\ref{EQ:afTheta}) and the regimes
$a(f)=\Omega((1+\epsilon_0)^{\sqrt{n}})$, we can also obtain the
following condition:
\begin{equation}
f=\Omega(n^{1/4}), \nonumber
\end{equation}
which means that if $f$ scales faster than $n^{1/4}$, then the
result in (\ref{EQ:Tnupper}) is satisfied.
\end{remark}


\section{Achievability Result} \label{SEC:Achievability}

In this section, we show that the considered transmission scheme,
commonly used in wireless radio networks, is order-optimal in
underwater networks. Under a regular network of unit node density,
the conventional MH transmission is used and its achievable
throughput scaling is analyzed to show its order optimality.

The nearest-neighbor MH routing protocol~\cite{GuptaKumar:00} will
be briefly described with a slight modification. The basic procedure
of the MH protocol under our extended regular network is as follows:
\begin{itemize}
\item Divide the network into square routing cells, each of which
has unit area.
\item Draw a line connecting a S--D pair. A source transmits a packet to its destination using the nodes in the adjacent cells passing through the line.
\item The full power is used, i.e., the transmit powers at each node
is $P$.
\end{itemize}

Instead of original (continuous) MH transmissions, a bursty
transmission
scheme~\cite{OzgurLevequeTse:07,OzgurJohariTseLeveque:10}, which
uses only a fraction $1/a(f)N(f)$ of the time for actual
transmission with instantaneous power $a(f)N(f)P$ per node, is used
to simply apply the analysis for networks with no power limitation
to our network model. With this scheme, the received signal power
from the desired transmitter, the noise psd, and the total
interference power from the set $I \subset \{1,\cdots,n\}$ have the
same scaling, i.e., $\Theta(N(f))$, and the (instantaneous) received
signal-to-interference-and-noise ratio (SINR) is kept at $\Theta(1)$
under the narrow-band model, which will be obviously shown in the
proof of Theorem~\ref{TH:achievable}.

The achievable rate of MH is now shown by quantifying the amount of
interference.

\begin{lemma}   \label{LEM:interference}
Suppose that a regular network of unit node density uses the MH
protocol with burstiness. Then, the total interference power from
other simultaneously transmitting nodes, corresponding to the set
$I\subset \{1,\cdots,n\}$, is upper-bounded by $\Theta(N(f))$, where
$N(f)$ denotes the psd of noise $n_i$ at receiver
$i\in\{1,\cdots,n\}$.
\end{lemma}

\begin{figure}[t!]
  \begin{center}
  \leavevmode \epsfxsize=0.60\textwidth   
  \leavevmode 
  \epsffile{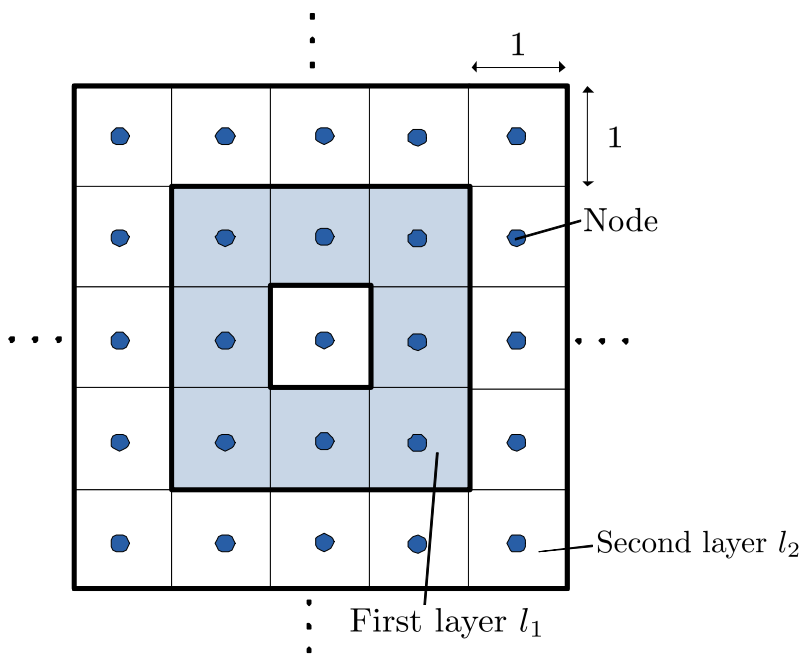}
  \caption{Grouping of interference routing cells in extended networks. The first layer $l_1$ represents the outer 8 shaded cells.}
  \label{FIG:layer}
  \end{center}
\end{figure}

\begin{proof}
There are $8k$ interfering routing cells, each of which includes one
node, in the $k$-th layer $l_k$ of the network as illustrated in
Fig.~\ref{FIG:layer}. Then from (\ref{EQ:af}), (\ref{EQ:signal}),
and (\ref{EQ:hki}), the total interference power at each node from
simultaneously transmitting nodes is upper-bounded by
\begin{eqnarray}
\sum_{k=1}^{\infty}(8k)\frac{a(f)N(f)P}{c_0k^{\alpha}a(f)^{k}}\!\!\!\!\!\!\!&&=\frac{8N(f)P}{c_0}\sum_{k=1}^{\infty}\frac{1}{k^{\alpha-1}a(f)^{k-1}}
\nonumber\\ &&\le
\frac{8N(f)P}{c_0}\sum_{k=1}^{\infty}\frac{1}{a(f)^{k-1}} \nonumber\\
&&\le c_6N(f), \nonumber
\end{eqnarray}
where $c_0$ and $c_6$ are some positive constants independent of
$n$, which completes the proof.
\end{proof}

Note that the signal power no longer decays polynomially but rather
exponentially with propagation distance in our network. This implies
that the absorption term $a(f)$ in (\ref{EQ:af}) will play an
important role in determining the performance. It is also seen that
the upper bound on the total interference power does not depend on
the spreading factor $\alpha$. Using Lemma~\ref{LEM:interference},
it is now possible to simply obtain a lower bound on the capacity
scaling in the network, and hence the following result presents the
achievable rates under the MH protocol.

\begin{theorem} \label{TH:achievable}
In an underwater regular network of unit node density,
\begin{equation}   \label{EQ:achievablerate}
T(n)=\Omega\left(\frac{n^{1/2}}{a(f)N(f)}\right)
\end{equation}
is achievable.
\end{theorem}

\begin{proof}
Suppose that only a fraction $1/a(f)N(f)$ of the time for actual
transmission is used under the MH protocol with burstiness. Then,
the SINR seen by any receiver is expressed as $\Omega(1)$ with an
instantaneous transmit power $a(f)N(f)P$ since the total
interference power is given by $O(N(f))$. Since the Gaussian is the
worst additive noise~\cite{Medard:00,DiggaviCover:01}, assuming it
lower-bounds the throughput. Hence, by assuming full CSI at the
receiver, the achievable throughput per S--D pair is lower-bounded
by
\begin{eqnarray}
&&\frac{1}{a(f)N(f)}\log (1+\text{SINR}) \nonumber\\
\ge\!\!\!\!\!\!\!&&\frac{1}{a(f)N(f)}\log\left(1+\frac{N(f)P/c_0}{N(f)+c_6N(f)}\right),
\label{EQ:SINR} \nonumber
\end{eqnarray}
for some positive constants $c_0$ and $c_6$ independent of $n$,
thereby providing the rate of
\begin{equation}
\Omega\left(\frac{1}{a(f)N(f)}\right). \nonumber
\end{equation}
Since the number of hops per S--D pair is given by $O(\sqrt{n})$,
there exist $\Omega(\sqrt{n})$ source nodes that can be active
simultaneously, and therefore the total throughput is finally given
by (\ref{EQ:achievablerate}), which completes the proof of the
theorem.
\end{proof}

Now it is examined how the upper bound shown in
Section~\ref{SEC:Upper} is close to the achievable throughput
scaling.

\begin{remark}
Based on Theorems~\ref{TH:Cutset} and~\ref{TH:achievable}, when
$a(f)=\Omega\left((1+\epsilon_0)^{\sqrt{n}}\right)$, i.e.,
$f=\Omega(n^{1/4})$, it is easy to see that the achievable rate and
the upper bound are of the same order up to $n^{\epsilon}$, where
$\epsilon$ and $\epsilon_0$ are vanishingly small positive
constants. MH is therefore order-optimal in regular networks with
unit node density under the above attenuation regimes.
\end{remark}

We also remark that applying the hierarchical cooperation
strategy~\cite{OzgurLevequeTse:07} may not be helpful to improve the
achievable throughput due to long-range MIMO transmissions, which
severely degrade performance in highly power-limited
networks.\footnote{In wireless radio networks of unit node density,
the hierarchical cooperation provides a near-optimal throughput
scaling for the operating regimes $2<\alpha<3$, where $\alpha$
denotes the path-loss exponent that is greater than
2~\cite{OzgurLevequeTse:07}. Note that the analysis
in~\cite{OzgurLevequeTse:07} is valid under the assumption that
$\alpha$ is kept at the same value on all levels of hierarchy.} To
be specific, at the top level of the hierarchy, the transmissions
between two clusters having distance $O(\sqrt{n})$ become a
bottleneck, and thus cause a significant throughput degradation. It
is further seen that even with the random phase model, which may
enable us to obtain enough degrees-of-freedom gain, the benefit of
randomness cannot be exploited because of the power limitation.


\section{Extension to Random Networks} \label{SEC:Random}

In this section, we would like to mention a random network
configuration, where $n$ S--D pairs are uniformly and independently
distributed on a square.

\begin{figure}[t!]
  \begin{center}
  \leavevmode \epsfxsize=0.43\textwidth   
  \leavevmode 
  \epsffile{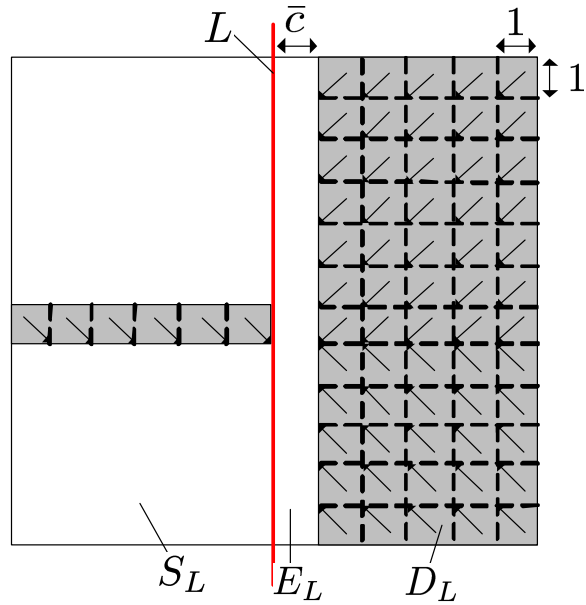}
  \caption{The node displacement to square vertices, indicated by arrows. The empty zone $E_L$ with width constant $\bar{c}$ is assumed for simplicity.}
  \label{FIG:displacement2}
  \end{center}
\end{figure}

We first discuss an upper bound for extended networks of unit node
density. A precise upper bound can be obtained using the binning
argument of~\cite{OzgurLevequeTse:07} (refer to Appendix V
in~\cite{OzgurLevequeTse:07} for the details). Consider the same cut
$L$, which divides the network area into two halves, as that in the
regular network case. For analytical convenience, we can
artificially assume the empty zone $E_L$, in which there are no
nodes in the network, consisting of a rectangular slab of width
$0<\bar{c}<\frac{1}{\sqrt{7}e^{1/4}}$, independent of $n$,
immediately to the right of the centerline (cut), as done
in~\cite{OzgurJohariTseLeveque:10} (see
Fig.~\ref{FIG:displacement2}).\footnote{ Although this assumption
does not hold in our random configuration, it is shown
in~\cite{OzgurJohariTseLeveque:10} that there exists a vertical cut
such that there are no nodes located closer than
$0<\bar{c}<\frac{1}{\sqrt{7}e^{1/4}}$ on both sides of this cut when
we allow a cut that is not necessarily linear. Such an existence is
proved by using percolation
theory~\cite{MeesterRoy:96,FranceschettiDouseTseThiran:07}. This
result can be directly applied to our network model since it only
relies on the node distribution but not the channel characteristics.
Hence, removing the assumption does not cause any change in
performance.} Let us state the following lemma.

\begin{lemma} \label{LEM:nodenum}
Assume a two dimensional extended network where $n$ nodes are
uniformly distributed. When the network area is divided into $n$
squares of unit area, there are fewer than $\log n$ nodes in each
square with high probability.
\end{lemma}

Since the result in Lemma~\ref{LEM:nodenum} depends on the node
distribution but not the channel characteristics, the proof
essentially follows that presented
in~\cite{FranceschettiDouseTseThiran:07}. By
Lemma~\ref{LEM:nodenum}, we now take into account the network
transformation resulting in a regular network with at most $\log n$
and $2\log n$ nodes, on the left and right, respectively, at each
square vertex except for the empty zone (see
Fig.~\ref{FIG:displacement2}). Then, the nodes in each square are
moved together onto one vertex of the corresponding square. More
specifically, under the cut $L$, the node displacement is performed
in the sense of decreasing the Euclidean distance between source
node $i\in S_L$ and the corresponding destination $k\in D_L$, as
shown in Fig.~\ref{FIG:displacement2}, which will provide an upper
bound on $d_L^{(i)}$ in (\ref{EQ:dL}). It is obviously seen that the
amount of power transfer under the transformed regular network is
greater than that under another regular network with at most $\log
n$ nodes at each vertex, located at integer lattice positions in a
square region of area $n$. Hence, the upper bound for random
networks is boosted by at least a logarithmic factor of $n$ compared
to that of regular networks discussed in Section~\ref{SEC:Upper}.

Now we turn our attention to showing an achievable throughput for
extended random networks. In this case, the nearest-neighbor MH
protocol~\cite{GuptaKumar:00} can also be utilized since our network
is highly power-limited. Then, the area of each routing cell needs
to scale with $2\log n$ to guarantee at least one node in a
cell~\cite{GuptaKumar:00,ElGamalMammenPrabhakarShah:06}.\footnote{When
methods from percolation theory are applied to our random
network~\cite{MeesterRoy:96,FranceschettiDouseTseThiran:07}, the
routing area constructed during the highway phase is a certain
positive constant that is less than 1 and independent of $n$. The
distance in the draining and delivery phases, corresponding to the
first and last hops of a packet transmission, is nevertheless given
by some constant times $\log n$, thereby limiting performance,
especially for the condition $a(f)=\omega(1)$. Hence, using the
protocol in~\cite{FranceschettiDouseTseThiran:07} indeed does not
perform better than the conventional MH case~\cite{GuptaKumar:00} in
random networks.} Each routing cell operates based on 9-time
division multiple access to avoid causing large interference to its
neighboring
cells~\cite{GuptaKumar:00,ElGamalMammenPrabhakarShah:06}. For the
routing with continuous MH transmissions (i.e., no burstiness),
since per-hop distance is given by $\Theta(\sqrt{\log n})$, the
received signal power from the intended transmitter and the SINR
seen by any receiver are expressed as
\begin{equation}
\frac{c_7P}{(\log n)^{\alpha/2}a(f)^{c_8\sqrt{\log n}}} \nonumber
\end{equation}
and
\begin{equation}
\Omega\left(\frac{1}{(\log n)^{\alpha/2}a(f)^{c_8\sqrt{\log
n}}N(f)}\right), \nonumber
\end{equation}
respectively, for some constants $c_7>0$ and $c_8\ge\sqrt{2}$
independent of $n$. Since the number of hops per S--D pair is given
by $O(\sqrt{n/\log n})$, there exist $\Omega(\sqrt{n/\log n})$
simultaneously active sources, and thus the total achievable
throughput $T(n)$ is finally given by
\begin{equation}
T(n)=\Omega\left(\frac{n^{1/2}}{(\log
n)^{(\alpha+1)/2}a(f)^{c_8\sqrt{\log n}}N(f)}\right) \nonumber
\end{equation}
for some constant $c_8\ge\sqrt{2}$ independent of $n$ (note that
this relies on the fact that $\log(1+x)$ can be approximated by $x$
for small $x>0$). Hence, using the MH protocol results in at least a
polynomial decrease in the throughput compared to the regular
network case shown in Section~\ref{SEC:Achievability}.\footnote{In
terrestrial radio channels, there is a logarithmic gap in the
achievable scaling laws between regular and random
networks~\cite{GuptaKumar:00,XieKumar:04}.} This comes from the fact
that the received signal power tends to be mainly limited due to
exponential attenuation with transmission distance
$\Theta(\sqrt{\log n})$. Note that in underwater networks,
randomness on the node distribution causes a huge performance
degradation on the throughput scaling. Therefore, we may conclude
that the existing MH scheme does not satisfy the order optimality
under extended random networks regardless of the attenuation
parameter $a(f)$.


\section{Conclusion} \label{SEC:Conc}

The attenuation parameter and the capacity scaling laws have been
characterized in a narrow-band channel of underwater acoustic
networks. Provided that the carrier frequency $f$ scales at
arbitrary rates relative to the number $n$ of nodes, the
information-theoretic upper bound and the achievable throughput were
derived as a function of the attenuation parameter $a(f)$ in
extended regular networks. Specifically, based on the power transfer
argument, the upper bound was shown to decrease in inverse
proportion to $a(f)$. In addition, to show the achievability result,
the nearest-neighbor MH protocol was introduced with a simple
modification, and its throughput scaling was analyzed. We proved
that the MH protocol is order-optimal as long as the frequency $f$
scales faster than $n^{1/4}$. Our scaling results were also extended
to the random network scenario, where it was shown that the
conventional MH scheme does not satisfy the order optimality for all
the operating regimes.

Suggestions for further research include, in dense networks of unit
area, analyzing an upper bound and designing an achievable scheme
whose throughput scaling is close to the upper bound.


\appendix

\section{Appendix}

\subsection{Proof of Lemma~\ref{LEM:dL}} \label{PF:dL}

\begin{figure}[t!]
  \begin{center}
  \leavevmode \epsfxsize=0.60\textwidth   
  \leavevmode 
  \epsffile{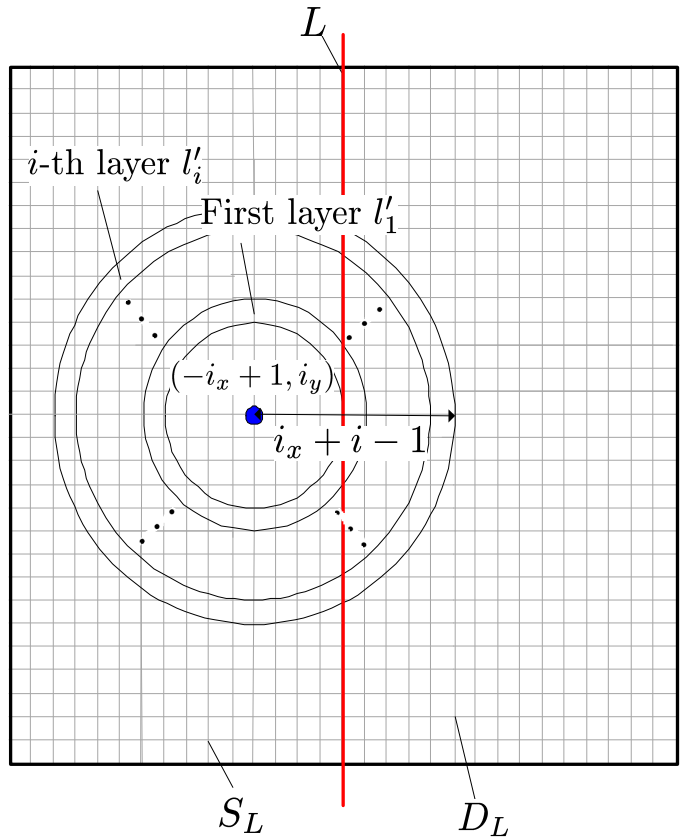}
  \caption{Grouping of destination nodes in extended networks. There exist $\Theta(i_x)$ nodes in the first layer $l_1'$. This figure indicates the case where one source is located at the position $(-i_x+1,i_y)$. The destination nodes are regularly placed at spacing 1 on the right half of the cut $L$.}
  \label{FIG:displacement}
  \end{center}
\end{figure}

Upper and lower bounds on $d_L^{(i)}$ can be found by using the
node-indexing and layering techniques similar to those shown in
Section VI of~\cite{ShinJeonDevroyeVuChungLeeTarokh:08}. As
illustrated in Fig.~\ref{FIG:displacement}, layers are introduced,
where the $i$-th layer $l_i'$ of the network represents the ring
with width 1 drawn based on a source node, whose coordinate is given
by $(-i_x+1,i_y)$, where $i\in\{1,\cdots,\sqrt{n}\}$. More
specifically, the ring is enclosed by the circumferences of two
circles, each of which has radius $i_x+i-1$ and $i_x+i-2$,
respectively, at its same center (see Fig.~\ref{FIG:displacement}).
We can see that there exist $\Theta(i_x+i)$ nodes in the layer
$l_i'$ since the area of $l_i'$ is given by $\pi(2i_x+2i-3)$. Then
from (\ref{EQ:dL}), the term $d_{L}^{(i)}$ is given by
\begin{equation}
d_{L}^{(i)}=
\frac{1}{c_0}\sum_{k_x=1}^{\sqrt{n}/2}\sum_{k_y=1}^{\sqrt{n}}\frac{1}{\left((i_x+k_x-1)^2+(i_y-k_y)^2\right)^{\alpha/2}a(f)^{\sqrt{(i_x+k_x-1)^2+(i_y-k_y)^2}}}.
\nonumber
\end{equation}
It is further assumed that all the nodes in each layer are moved
onto the innermost boundary of the corresponding ring, which
provides an upper bound for $d_{L}^{(i)}$. Since there is no node,
located on the right half of the cut $L$, in the first layer $l_1'$,
$d_{L}^{(i)}$ is then upper-bounded by
\begin{eqnarray} \label{EQ:dLupp}
d_{L}^{(i)}\!\!\!\!\!\!\!
&&\le\frac{1}{c_0}\sum_{k'=i_x}^{\infty}\frac{c_9k'}{k'^{\alpha}a(f)^{k'}}
\nonumber\\
&&\le\frac{c_9}{c_0i_x^{\alpha-1}}\sum_{k'=i_x}^{\infty}\frac{1}{a(f)^{k'}}
\nonumber\\ &&\le
\frac{c_9}{c_0i_x^{\alpha-1}}\left(\frac{1}{a(f)^{i_x}}+\int_{i_x}^{\infty}\frac{1}{a(f)^x}dx\right) \nonumber\\
&& \le\frac{c_{10}}{i_x^{\alpha-1}a(f)^{i_x}}, \nonumber
\end{eqnarray}
where $c_0$, $c_9$, and $c_{10}$ are some positive constants
independent of $n$. Here, the fourth inequality holds since
$a(f)>1$. To get a lower bound for $d_{L}^{(i)}$, nodes in each
layer are now moved onto the outermost boundary of the corresponding
ring. Let $\delta_i$ denote the fraction of nodes that are placed on
the right half of the network among nodes in the $i$-th layer
$l_i'$, which is obviously independent of $n$ irrespective of
$i\in\{1,\cdots,\sqrt{n}/2\}$. Thus, the lower bound similarly
follows
\begin{eqnarray} \label{EQ:dLlow}
d_{L}^{(i)}\!\!\!\!\!\!\!&&
\ge\frac{1}{c_0}\sum_{k'=i_x}^{i_x+\sqrt{n}/2-1}\frac{c_9\delta_{k'-i_x+1}k'}{k'^{\alpha}a(f)^{k'}}
\nonumber\\
&&\ge\frac{c_9\min\{\delta_1,\cdots,\delta_{\sqrt{n}/2}\}}{c_0}\sum_{k'=i_x}^{i_x+\sqrt{n}/2-1}\frac{1}{k'^{\alpha-1}a(f)^{k'}}
\nonumber\\&&\ge \frac{c_{11}}{i_x^{\alpha-1}a(f)^{i_x}}, \nonumber
\end{eqnarray}
where $c_0$, $c_9$, and $c_{11}$ are some positive constants
independent of $n$, which finally yields (\ref{EQ:dLupper}). This
completes the proof.


\subsection{Proof of Lemma~\ref{LEM:singular}} \label{PF:singular}

Since the size of matrix ${\bf F}_{L}$ is given by
$\Theta(n)\times\Theta(n)$, the analysis essentially follows the
argument in~\cite{OzgurLevequeTse:07} with a slight modification
(refer to Appendix III in~\cite{OzgurLevequeTse:07} for more precise
description). Suppose that
$a(f)=\Omega\left((1+\epsilon_0)^{\sqrt{n}}\right)$ for an
arbitrarily small $\epsilon_0>0$. Then in the following, from the
result of Lemma~\ref{LEM:dL}, we derive $\sum_{k\in
D_{L}}\left|[{\bf F}_{L}]_{ki}\right|^2$ and an upper bound for
$\sum_{i\in S_{L}}\left|[{\bf F}_{L}]_{ki}\right|^2$:
\begin{eqnarray} \label{EQ:FL2DL2}
\underset{k\in D_{L}}\sum \left|\left[{\bf
F}_{L}\right]_{ki}\right|^2\!\!\!\!\!\!\!&&=\underset{k\in
D_{L}}\sum \left|\frac{1}{\sqrt{d_{L}^{(i)}}}\left[{\bf
H}_{L}\right]_{ki}\right|^2 \nonumber\\ &&=\frac{\sum_{k\in D_{L}}
\left|\left[{\bf H}_{L}\right]_{ki}\right|^2}{\sum_{k\in
D_{L}}A(r_{ki},f)^{-1}} \nonumber\\ &&=1, \nonumber
\end{eqnarray}
where the second equality comes from (\ref{EQ:af}), (\ref{EQ:hki}),
and (\ref{EQ:dL}), and
\begin{eqnarray}
\underset{i\in S_{L}}\sum \left|\left[{\bf
F}_{L}\right]_{ki}\right|^2 \!\!\!\!\!\!\!&&=\underset{i\in
S_{L}}\sum \left|\frac{1}{\sqrt{d_{L}^{(i)}}}\left[{\bf
H}_{L}\right]_{ki}\right|^2 \nonumber\\ &&\le
c_0\sum_{i_x=1}^{\sqrt{n}/2}\sum_{i_y=1}^{\sqrt{n}}\frac{i_x^{\alpha-1}a(f)^{i_x}}{\left((i_x+k_x-1)^2+(i_y-k_y)^2\right)^{\alpha/2}a(f)^{\sqrt{(i_x+k_x-1)^2+(i_y-k_y)^2}}}
\nonumber\\ &&\le
c_0\sum_{i_x=1}^{\sqrt{n}/2}\sum_{i_y=1}^{\sqrt{n}}\frac{a(f)^{i_x}}{\sqrt{(i_x+k_x-1)^2+(i_y-k_y)^2}a(f)^{\sqrt{(i_x+k_x-1)^2+(i_y-k_y)^2}}}
\nonumber\\
&&\le
c_0\sum_{i_x=1}^{\sqrt{n}/2}\frac{1}{i_x+k_x-1}\left(\sum_{i_y=1}^{\sqrt{n}}\frac{1}{a(f)^{\sqrt{i_x^2+i_y^2}-i_x}}\right)
\nonumber\\ &&\le
c_0\sum_{i_x=1}^{\sqrt{n}/2}\frac{1}{i_x}\left(\sum_{i_y=1}^{\sqrt{n}}\frac{1}{a(f)^{i_y^2/\left(\sqrt{i_x^2+i_y^2}+i_x\right)}}\right)
\nonumber\\ &&\le
c_0\sum_{i_x=1}^{\sqrt{n}/2}\frac{1}{i_x}\left(\sum_{i_y=1}^{\sqrt{n}}\frac{1}{(1+\epsilon_0)^{c_{12}i_y^2}}\right)\nonumber\\
&&\le
c_0\left(1+\int_1^{\sqrt{n}/2}\frac{1}{x}dx\right)\left(\frac{1}{(1+\epsilon_0)^{c_{12}}}+\int_1^{\sqrt{n}}\frac{1}{(1+\epsilon_0)^{c_{12}y}}dy\right)
\nonumber\\ &&\le c_{13} \log n \nonumber
\end{eqnarray}
for some positive constants $c_{12}$ and $c_{13}$ independent of
$n$, where the second inequality holds since $\alpha=1$ provides the
highest upper bound for all $1\le\alpha\le2$. The fifth inequality
comes from the fact that
$a(f)=\Omega\left((1+\epsilon_0)^{\sqrt{n}}\right)$. Hence, it is
proved that both scaling results are the same as the regular network
case shown in~\cite{OzgurLevequeTse:07}.

Now we are ready to prove the inequality in (\ref{EQ:EFL2}).
Following the same line as that in Appendix III
of~\cite{OzgurLevequeTse:07}, we thus have
\begin{equation}
E\left[\tr\left(\left({\bf F}_{L}^{H}{\bf
F}_{L}\right)^q\right)\right]\le C_q n(c_{14}\log n)^{q}, \nonumber
\end{equation}
where $C_q=\frac{(2q)!}{q!(q+1)!}$ is the Catalan number for any $q$
and $c_{14}>0$ is some constant independent of $n$. Then, from the
property $\|{\bf F}_{L}\|_2^2=\lim_{q\rightarrow\infty}\tr(({\bf
F}_{L}^{H}{\bf F}_{L})^q)^{1/q}$ (refer to~\cite{HornJohnson:99}),
the expectation of the term $\|{\bf F}_{L}\|_2^2$ is upper-bounded
by
\begin{eqnarray}
E\left[\left\|{\bf
F}_{L}\right\|_2^2\right]\!\!\!\!\!\!\!&&\le\lim_{q\rightarrow\infty}\left(E\left[\tr\left(\left({\bf
F}_{L}^{H}{\bf F}_{L}\right)^q\right)\right]\right)^{1/q} \nonumber\\
&&\le\lim_{q\rightarrow\infty}\left(C_q
n\left(c_{14}\log n\right)^{q}\right)^{1/q} \nonumber\\
&&=4c_{14}\log n, \nonumber
\end{eqnarray}
where the equality holds since
$\lim_{q\rightarrow\infty}C_q^{1/q}=4$. Here, the first inequality
comes from dominated convergence theorem and Jensen's inequality.
This completes the proof.


\newpage


\end{document}